\documentclass[%
twocolumn,
superscriptaddress,
longbibliography,
amsmath,amssymb,
aps,
prb,
]{revtex4-2}
\setcitestyle{numbers,square}

\usepackage{braket}
\usepackage{float}
\usepackage{placeins}
\usepackage{graphicx}
\usepackage{dcolumn}
\usepackage{bm}
\usepackage[normalem]{ulem} 
\usepackage{xcolor}
\usepackage{tabularx,booktabs}

\usepackage{hyperref}
\hypersetup{
    colorlinks,%
    citecolor=blue,%
    linkcolor=blue,%
    urlcolor=blue
}

\newcommand{\orcid}[1]{\href{https://orcid.org/#1}{\includegraphics[width=8pt]{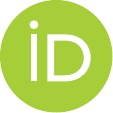}}}

\begin{document}

\title{Interacting Virtual Topological Phases in Defect-Rich 2D Materials}

\author{F. Crasto de Lima\orcid{0000-0002-2937-2620}} 
\email{felipe.lima@ilum.cnpem.br}
\affiliation{Ilum School of Science, Brazilian Center for Research in Energy and Materials (CNPEM), Zip Code 13083-970, Campinas, São Paulo, Brazil.}

\author{Roberto H. Miwa\orcid{0000-0002-1237-1525}} 
\email{hiroki@ufu.br}
\affiliation{Instituto de Física, Universidade Federal de Uberlândia, C.P. 593, Uberlândia, MG 38400-902, Brazil}

\author{Caio Lewenkopf\orcid{0000-0002-2053-2798}} 
\affiliation{Instituto de F\'isica, Universidade Federal do Rio de Janeiro, 21941-972 Rio de Janeiro, Rio de Janeiro, Brazil}

\author{Adalberto Fazzio\orcid{0000-0001-5384-7676}} 
\email{adalberto.fazzio@ilum.cnpem.br}
\affiliation{Ilum School of Science, Brazilian Center for Research in Energy and Materials (CNPEM), Zip Code 13083-970, Campinas, São Paulo, Brazil.}

\date{\today}

\begin{abstract}
We investigate the robustness of {\it virtual} topological states -- topological phases away from the Fermi energy -- against the electron-electron interaction and band filling. 
As a case study, we employ a realistic model to investigate the properties of vacancy-driven topological phases in transition metal dichalcogenides (TMDs) and establish a connection between the degree of localization of topological wave functions, the vacancy density, and the electron-electron interaction strength with the topological phase robustness. 
We demonstrate that electron-electron interactions play a crucial role in degrading topological phases thereby determining the validity of single-particle approximations for topological insulator phases.
Our findings can be naturally extended to {\it virtual} topological phases of a wide range of materials. 
\end{abstract}

\maketitle

\section{Introduction}

Topological insulators (TIs) are a new class of materials that exhibit insulating bulk behavior, but conduct electricity on their surfaces \cite{PRLkane2005, SCIENCEbernevig2006, PRLfu2007}, due to the presence of topologically protected metallic surface states that are robust against disorder \cite{Hasan2010, Qi2011, Moessner2021}.
This unique property, a hallmark of TIs, combined with an unusual spin texture of their topological surface states, has attracted enormous attention in the condensed matter physics and materials science \cite{Breunig2022}.
Remarkably, while theory predicts thousands of materials exhibiting topological phases \cite{Zhang2019, Vergniory2019, Tang2019}, the number of experimental realizations is still surprisingly small.
This puzzling discrepancy can be explained, in part, by noting that approximately 88\% of the predicted topological phases correspond to {\it virtual} electronic gaps, here referred to as energy gaps that are energetically far from the Fermi level \cite{Zhang2019, Vergniory2019, Tang2019,SCIENCEvergniory2022, PRXkruthoff2017, NATPHYSslager2012}, requiring an electronic doping that hinders their experimental observation.
However, a more fundamental challenge lies in understanding the influence of charge carrier doping and the electron-electron interaction on the electronic structure and topological properties of these materials.

For simplicity, let us focus on 2D materials.
Here, as in the 3D case, numerous works have been dedicated to the study of topological-trivial/non-trivial transitions that occur as a function of temperature \cite{PRLmonserrat2016, PRBfocassio2021, PRBchen2024}, external fields \cite{PNASkim2011, NLliu2015}, disorder \cite{Assuncao2024, Regis2024}, and alloying \cite{PRBchang2016, PRMzhang2020, 2DMlima2023, PRBmota2024}. 
However, despite the diverse range of mechanisms investigated for inducing phase transitions, the topological phase robustness with electron doping to access virtual topological states remains relatively unexplored.
In this context, electron-electron repulsion plays a crucial role, particularly depending on the features of the electronic states associated with the non-trivial gap.
The competition between topological phases and interaction induced strong electronic correlations  has been studied for fixed electron number within the Kane-Mele-Hubbard model \cite{PRBrachel2010, PRLhohenadler2011}.
Our focus, instead, lies on the topological systems predicted in Refs.~\cite{Zhang2019, Vergniory2019, Tang2019} which typically do not exhibit strong electronic correlations and can be treated within a mean-field approach.

In summary, notwithstanding the significant progress \cite{Rachel2018}, the effect of electron-electron repulsion on topological gaps away from the Fermi level remains unexplored. 

This paper addresses the critical question of how electron filling influences the stability of predicted virtual topological phase transitions, revealing the conditions under which these transitions remain robust and the associated changes in the electronic band structure.

\begin{figure}[t]
    \includegraphics[width=1\linewidth]{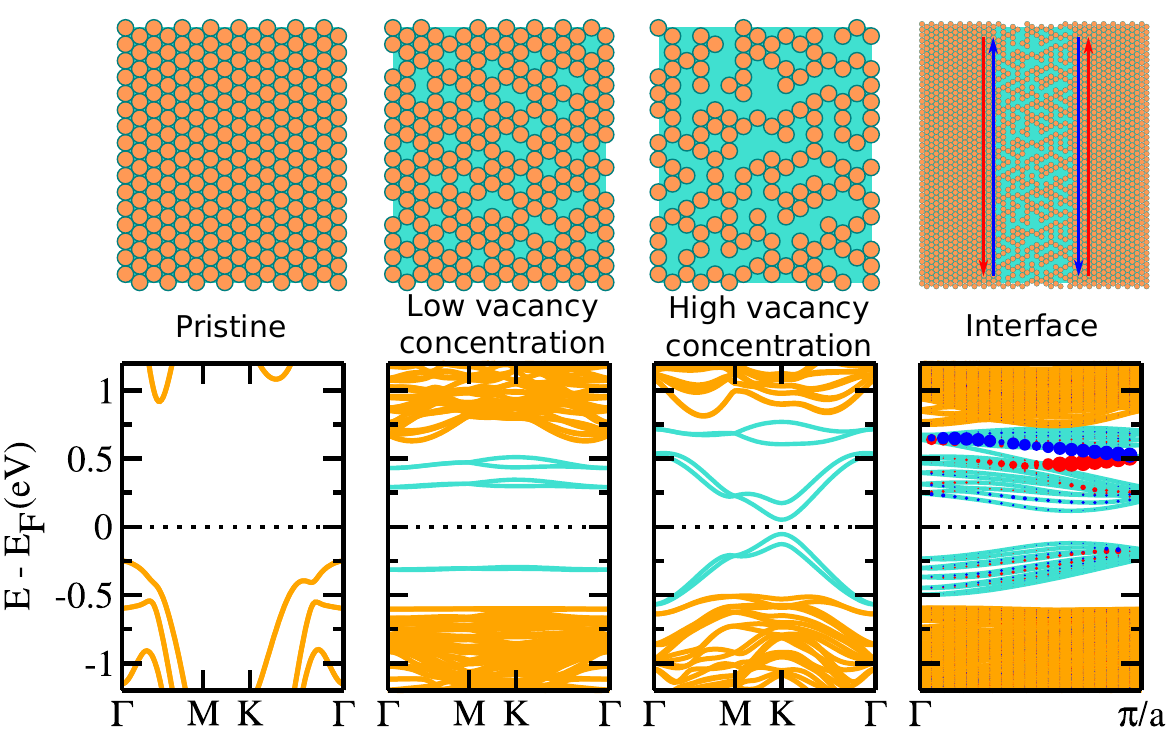}
    \caption{\label{fig:schema} 
    Vacancy-induced topological phase in 2D MX$_2$: at critical 
    chalcogen vacancy concentration, a topological gap forms, yielding spin-polarized edge states on unoccupied levels. The band structures are DFT calculations for PtSe$_2$; dashed lines mark the Fermi level; blue/red circles denote spin-up/down edge states.}
\end{figure}

An ideal system for studying these features should allow the manipulation of wave function localization to increase electron-electron repulsion. 
Surprisingly, localized vacancy states in 2D-TMDs exhibit virtual non-trivial topological states that emerge beyond a critical vacancy concentration \cite{PRBlima2017, NLlima2021}, as shown on Fig.~\ref{fig:schema}. 
With increasing vacancy concentration (corresponding to panels from left to right), vacancy states emerge, resulting in the formation of midgap states.
These midgap states, beyond a critical concentration, give rise to topological states observable at the system boundaries.  
A detailed discussion is presented below.
Additionally, given the localized nature of such defect states, electron-electron repulsion can compete both with the vacancy level splittings \cite{PRBabsor2017} and the formation of a topological phase \cite{NLlima2021}.
Our study explores the electron-electron repulsion in recent predicted non-trivial topological states in 2D TMDs.
Through {\it ab initio} calculations, we demonstrate the existence of chalcogen vacancy-induced defect states in 1T-MX$_2$ systems, where M=Ni, Pd, Pt and X=S, Se, Te.
By constructing a general minimal model within a Hubbard mean-field tight-binding approach, we validate the non-trivial/trivial phase transition in these systems as a function of the vacancy concentration and electron-electron repulsion. 

Our calculations reveal the formation of three bands within the gap for the semiconducting systems, see Fig.~\ref{fig:schema}, originating from the $d$-states of the transition metal atoms adjacent to the vacant site \cite{PRMfreire2022}.
Given the absence of the chalcogen atom, two electrons occupy these vacancy states, resulting in a $\nu=1/3$ filling for neutral systems.

Notably, our calculations for PtSe$_2$ indicate that vacancies concentrations exceeding a critical concentration of $2.5\times 10^{14}$\,cm$ ^{-2}$
\cite{NATMATlin2017} induce gap opening between the forth and fifth bands (at $\nu=2/3$ filling), that are characterized as non-trivial.
Since these bands are unoccupied in charge-neutral systems, this constitutes an example of a {\it virtual topological phase}. 
We employ this system as a showcase to study the influence of electron-electron repulsion on the virtual topological gap and its robustness upon filling.

\section{Methods}

\subsection{Ab initio calculations}

We perform density functional theory (DFT) calculations for the MX$_2$ system within a plane-wave base code, Vienna ab initio simulation package (VASP) \cite{PRBkresse1993, PRBkresse1996}, with cut-off energy for the plane wave expansion of $400$\,eV.
We employ
the PBE \cite{PRLperdew1997} functional to describe the exchange and correlation terms, solving the system in a uniform grid of $11 \times 11 \times 1$ k-points for the unit cell, and using the same density for any supercells.
All atomic positions are relaxed until forces converged below
$10^{-3}$\,eV/{\AA}, with the electron-ion interaction described by the projected augmented wave method \cite{PRBblochl1994, PRBkresse1999}.

\subsection{Model Hamiltonian}

A minimal model to explore an insulator with virtual topological states requires at least two energy gaps, one trivial and the other non-trivial.
A three orbital model is sufficient to capture the essential features of virtual topology.
Specifically, for chalcogen vacancies in MX$_2$ systems, a three orbital model accurately describes the states close to the Fermi energy.
We consider such three orbitals model, incorporating intra- and inter-vacancy orbital interactions, spin-orbit effects, and on-site Hubbard electron-electron repulsion
\begin{equation}
    H = H_{\rm intra} + H_{\rm inter} + H_{\rm U}.
\end{equation}

The intra-vacancy term couples the three Pt orbitals surrounding
a vacancy
\begin{equation*}
    H_{\rm intra} = \sum_{\langle ij \rangle}  {\bf c}_i^{\dagger} (t \sigma_0 + i \lambda \nu_{ij} \sigma_z )  {\bf c}^{}_j,
\end{equation*}
where $t$ is the hopping amplitude, $\langle ij \rangle$ restricts the sum over first neighbors orbitals, $\lambda$ gives the SOC strength, $\nu_{ij}$ captures the sign of the SOC term \cite{PRLkane2005, NLlima2021}, ${\bf c}_i = \left[ c_{i \uparrow},\,c_{i \downarrow} \right]^{T}$ is the annihilation operator of electron at the site/orbital $i$ with spin projection $\uparrow\!\!/\!\!\downarrow$, while $\sigma_{0}$ and $\sigma_z$ are Pauli matrices.

The inter-vacancy term couples orbitals from adjacent vacancies
\begin{equation*}
    H_{\rm inter} = \sum_{\langle \langle ij \rangle \rangle } {\bf c}_i^{\dagger} (t' \sigma_0 + \lambda'\nu_{ij} \sigma_z )   {\bf c}^{}_j,
\end{equation*}
with the second neighbor hopping and SOC strengths given by $t' = \alpha t$ and $\lambda' = \beta \lambda$, with $0 \leq \alpha,\,\beta \leq 1$. 
$\alpha$ and $\beta$  parameterize vacancy concentration $n_{\rm vac}$ and their spatial localization. 
The low $n_{\rm vac}$ and high $n_{\rm vac}$ limits correspond to $\alpha,\beta \ll 1$ and $\alpha,\beta \approx 1$, respectively.
We consider here previously studied band topological systems \cite{Zhang2019, Vergniory2019, Tang2019} amenable to a mean-field description, as they typically do not exhibit strong electronic correlations.
We emphasize that DFT calculations for charged confined systems pose challenges associated with the background jellium \cite{PRLchagas2021, PRLshang2025}, providing additional justification for the inclusion of a Hubbard term in our approach.
The mean-field Hubbard term reads
\begin{equation*}
    H_{U} = U \sum_{i}  \left( n_{i \uparrow} \langle n_{i \downarrow} \rangle + n_{i \downarrow} \langle n_{i \uparrow} \rangle - \langle n_{i \uparrow} \rangle \langle n_{i \downarrow} \rangle \right), 
\end{equation*}
where $n_{i \sigma} = c_{i \sigma}^\dagger c_{i \sigma}^{}$ is the number operator and $U$ the on-site electronic interaction energy.
For finite $U$, the electronic problem is solved self-consistently.

The effective low energy electronic Hamiltonian matrix, expressed in the basis
$\{ {|1,\uparrow\rangle}, {|2,\uparrow\rangle}, |3,\uparrow\rangle , |1,\downarrow \rangle , |2,\downarrow\rangle, |3,\downarrow\rangle \}$, where the first index denotes one of the Pt-lone pair sites and the second the spin subspace, reads
\begin{widetext}
\begin{equation*}
    H= \begin{pmatrix}
        U( \langle n_{1\uparrow} \rangle - \eta)  & A_+({\bf k}) & B_-({\bf k}) & 0 & 0 & 0 \\
        A_+^*({\bf k}) & U (\langle n_{2\uparrow} \rangle - \eta) & C_+({\bf k})  & 0 & 0 & 0 \\
        B_-^*({\bf k}) & C_+^*({\bf k})  & U (\langle n_{3\uparrow}\rangle - \eta)& 0 & 0 & 0  \\
        0 & 0 & 0 & U (\langle n_{1\downarrow}\rangle - \eta) & A_-({\bf k}) & B_+({\bf k}) \\
         0 & 0 & 0 & A_-^*({\bf k}) & U ( \langle  n_{2\downarrow} \rangle - \eta ) & C_-({\bf k}) \\
        0 & 0 & 0 & B_+^*({\bf k}) & C_-^*({\bf k})  & U ( \langle n_{3\downarrow} \rangle - \eta ),
    \end{pmatrix}
\end{equation*}
\end{widetext}
with $\langle n_{i\gamma} \rangle $ the average number operator for the $i$-site and spin $\gamma$, $\displaystyle \eta = \sum_i \langle n_{i\uparrow} \rangle \langle n_{i\downarrow} \rangle $, and
\begin{eqnarray*}
    A_{\pm}({\bf k}) &=& t [1 + \alpha f_3({\bf k}) ] \pm  i \lambda [ 1 + \beta f_3({\bf k}) ]  \\
    B_{\pm}({\bf k}) &=& t[1+\alpha f_2({\bf k})] \pm i \lambda [1+\beta f_2({\bf k})]  \\
    C_{\pm}({\bf k}) &=& t[1+\alpha f_1({\bf k})] \pm i\lambda [1+\beta f_1({\bf k})].
\end{eqnarray*}
where $t$ and $\lambda$ are the hopping and intra-vacancy spin-orbit strength, respectively.
The inter-vacancy hopping and spin-orbit strength are lower than the intra-vacancy by the factors $\alpha$ and $\beta$ respectively.
The ${\bf k}$ dependent functions $f_j$ is associated with the periodic distribution of the vacancies, for a triangular distribution $f_1 = e^{i {\bf k}\cdot {\bf A}_1}$, $f_2 = e^{i {\bf k}\cdot {\bf A}_2}$, and $f_3 = e^{i {\bf k}\cdot ({\bf A}_2 - {\bf A}_1)}$; where ${\bf A}_i$ are the lattice vectors.
It is worth pointing out that we have considered other supercell geometries, as well as random vacancy distribution, which do not change our interpretation over the topological phases. {Additionally, while magnetic phases can emerge due to transition metal (TM) vacancies \cite{NATCOMavsar2020}, these defects are significantly less favorable energetically -- by about $5$\,eV --than the chalcogen vacancies we focus on \cite{PRMfreire2022}. Furthermore, our model calculations within a moderate interaction regime ($U \leq t$) do not yield any magnetic ground state for even electron fillings. Thus, magnetic or antiferromagnetic ordering is not expected to play a role in the systems and interaction range considered in this work.
}

\section{Results}

\begin{figure}[h!]
    \includegraphics[width=1\linewidth]{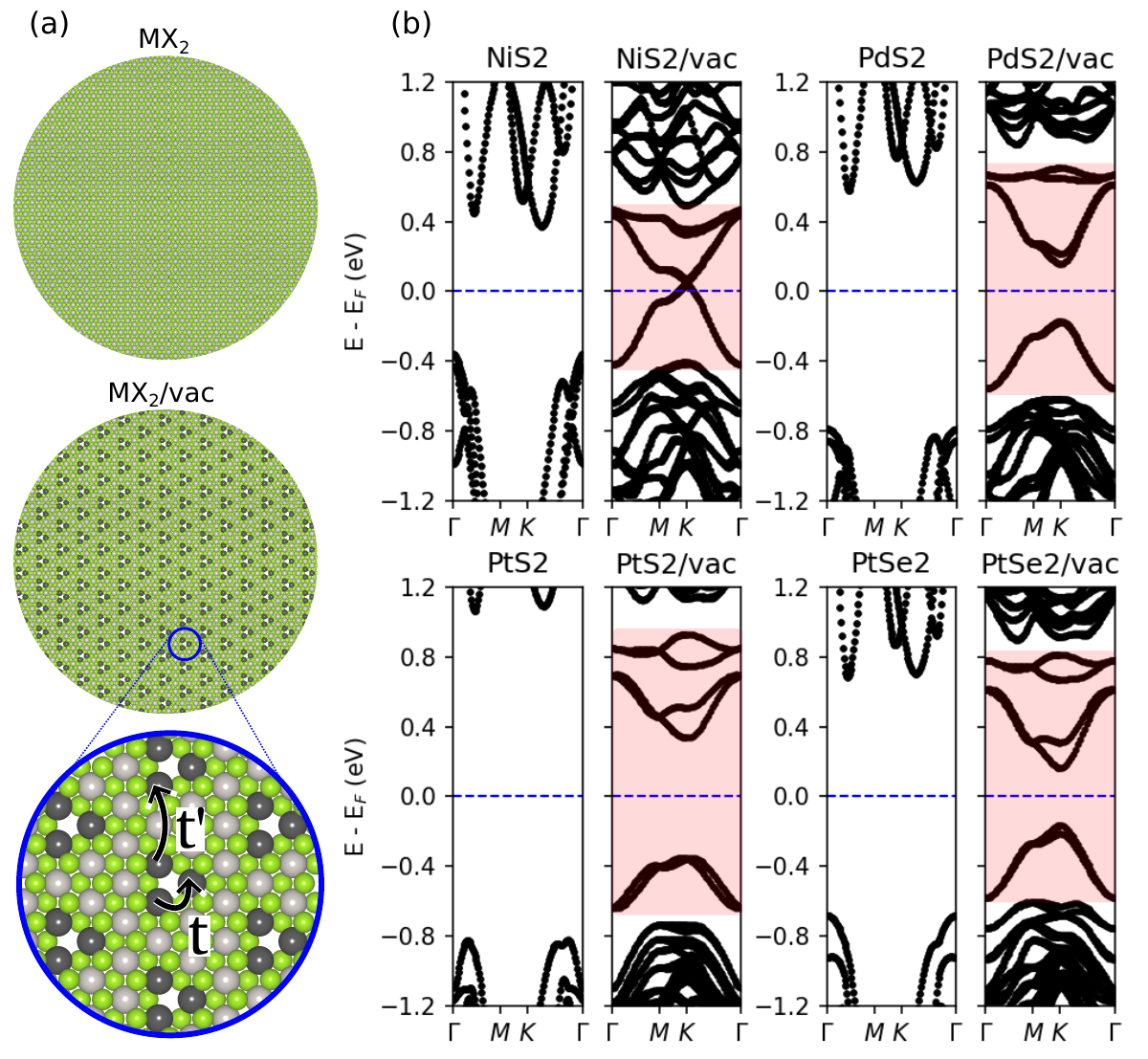}
    \caption{\label{fig:SM-1} 
    (a) Atomic structure of chalcogen vacancy on MX$_2$ transition metal dichalcogenides, showing the neighboring sites/vacancy hopping. (b) DFT simulation of the band structure for pristine MX$_2$ and MX$_2$ with $5\%$ of X atoms vacant, for the systems presenting isolated vacancy bands (shaded region).
    }
\end{figure}

\subsection{Electronic structure}

Figure~\ref{fig:SM-1} shows our ab initio results for the MX$_2$ systems, both pristine and with chalcogen vacancies.
All pristine systems with the lighter chalcogens (S and Se) are semiconducting. 
In contrast, MTe$_2$ exhibits metallic behavior for Ni and Pd, but remains semiconducting for the Pt system. 
We investigate the impact of chalcogen vacancies on the band structure, using a high concentration of approximately $ 10^{14}$\,{cm$^{-2}$}, consistent with experimentally achieved values in PtSe$_2$ \cite{NATMATlin2017} and MoS$_2$ \cite{NATMATli2015} systems.  
The band structures reveal the emergence of three midgap bands in the semiconducting systems.  
These states originate from the $d$-orbitals of the transition metal atoms neighboring the vacancy site \cite{PRMfreire2022}.

\begin{figure}[h!]
    \includegraphics[width=1\linewidth]{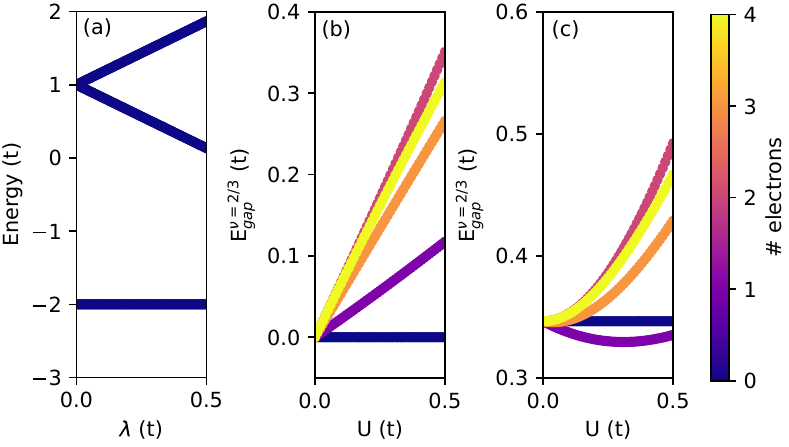}
    \caption{\label{fig:iso-electronic} Isolated vacancy electronic characteristics predicted by the model. (a) Intrinsic SOC effect on the $U=0$ system. Energy gap between the higher energy states as a function of $U$ (b) for $\lambda=0$, and (c) for $\lambda=0.1\,t$. The color bar is the same for all plots, indicating the number of electrons in the system.}
\end{figure}

\subsection{Non-trivial topology}

Let us begin by discussing the simplest case where there is no inter-vacancy coupling, namely,
$\alpha = \beta = 0$. 
In this limit the system is $k$-independent, and the wave functions are localized within individual vacancies.
Taking the first neighbor hopping as the energy scale $t=-1$, we have the electronic level picture shown in Fig.~\ref{fig:iso-electronic}.
Figure~\ref{fig:iso-electronic} shows the electronic levels for different scenarios taking  the first neighbor hopping as the energy scale, $t=-1$.
In the absence of SOC ($\lambda = 0$) and electron-electron repulsion ($U=0$), the low energy band is doubly degenerate and the higher-energy states exhibit 4-fold degeneracy.
SOC lifts the 4-fold degeneracy, opening a gap that scales linearly with the SOC strength.
Similarly, in the absence of SOC, the Hubbard term breaks the degeneracy between the topmost states, opening a gap that scales linearly with $U$. 
However, when both SOC and the Hubbard term are present, the gap between the top-most states scales with $U$ has $E_{\rm gap} \approx \sqrt{a U^2 + \lambda^2}$, where the coefficient $a$ depends on the electron filling, Fig.~\ref{fig:iso-electronic}(c). 

We analyze the system topological properties using the spin Chern number ${\cal C}_s = {\cal C}_{\uparrow} - {\cal C}_{\downarrow}$,
\begin{equation}
    {\cal C}_\gamma = \sum_n^{\rm occup} \oint_C {\bf A}_n^{(\gamma)} \cdot d {\bf k}
\end{equation}
where $\gamma=\uparrow,\downarrow$ and the Berry connection $ {\bf A}_n^{(\gamma)} = i \langle n,\,{\bf k}, \gamma | {\bm \nabla}_{\bf k} |n,\,{\bf k},\,\gamma \rangle$. 
Fully localized wave functions are eigenfunctions of the position operator $\hat{\bf x} = i {\bm \nabla}_{\bf k}$, leading to a trivial topology.

\begin{figure*}[t]
    \centering
    \includegraphics[width=0.85\linewidth]{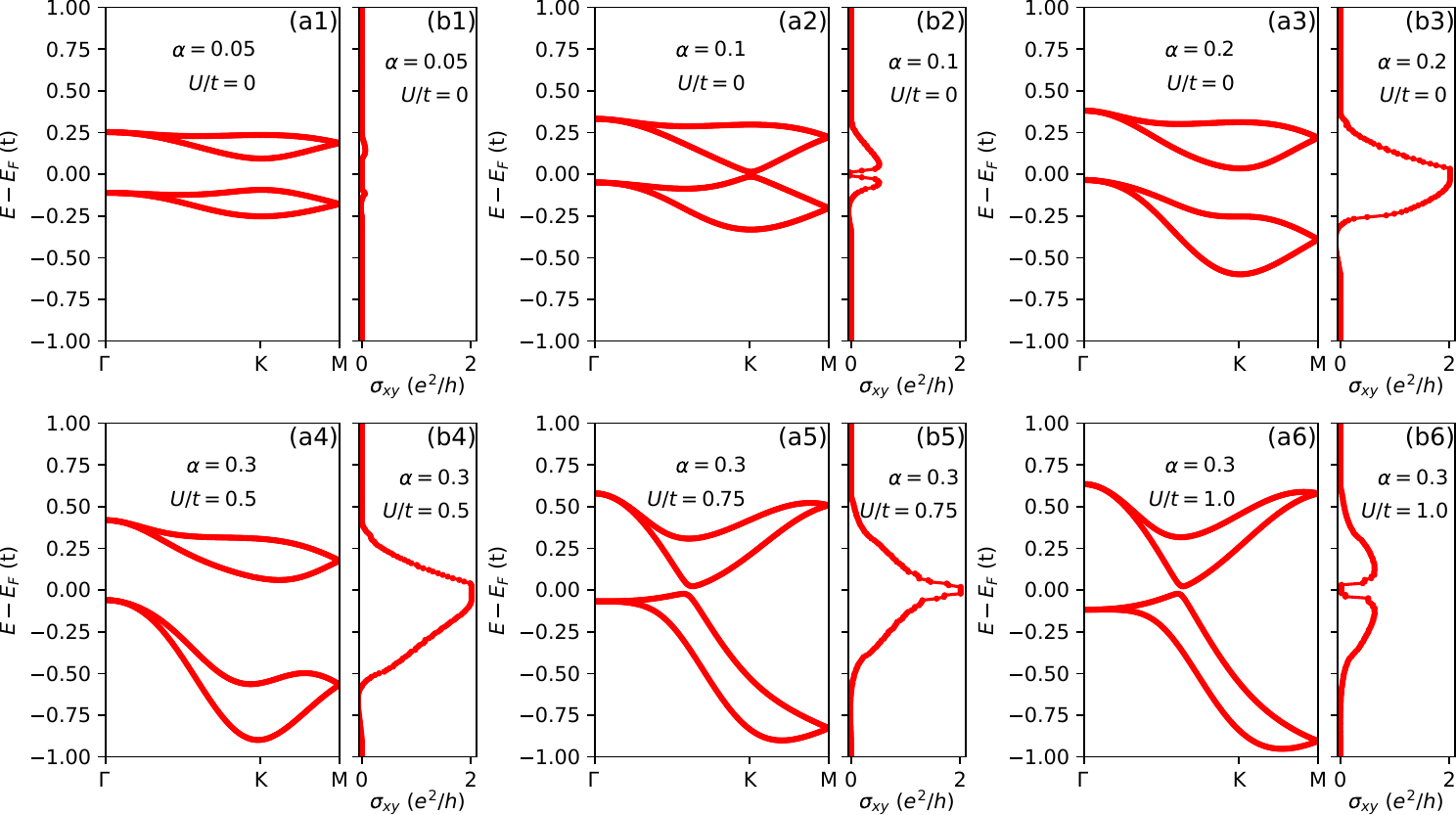}
    \caption{\label{fig:band-shc} 
    Model system with vacancies distributed on a hexagonal supercell. (a) Band structure along high-symmetry direction and (b) spin Hall conductivity $\sigma_{xy}$ for $\lambda=0.1t$, and $\beta=0$. Upper row: $U=0$ for selected values of $\alpha$. Lower row: $\alpha=0.3$ and $\nu=2/3$ filling for different values of $U$.}
\end{figure*}

\begin{figure}
    \includegraphics[width=0.9\linewidth]{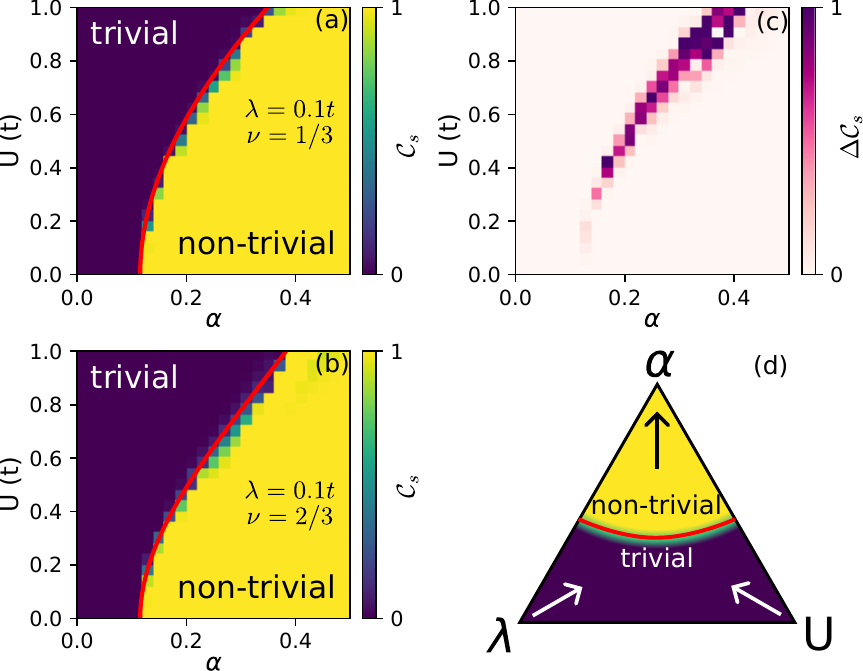}
    \caption{\label{fig:phase} 
    Topological phase diagram of the topmost energy gap for $U$ versus $\alpha$ with $\lambda=0.1\,t$ for (a) $\nu=1/3$ and (b) $\nu=2/3$. The red lines indicate the topological transition regions where $E_{\rm gap} \rightarrow 0$.
    (c) shows the difference between the phase diagrams (a) and (b).
    (d) Schematic phase diagram for the competition between vacancy density, SOC strength, and electron-electron interaction.
    }
\end{figure}

\begin{figure*}
    \includegraphics[width=0.85\linewidth]{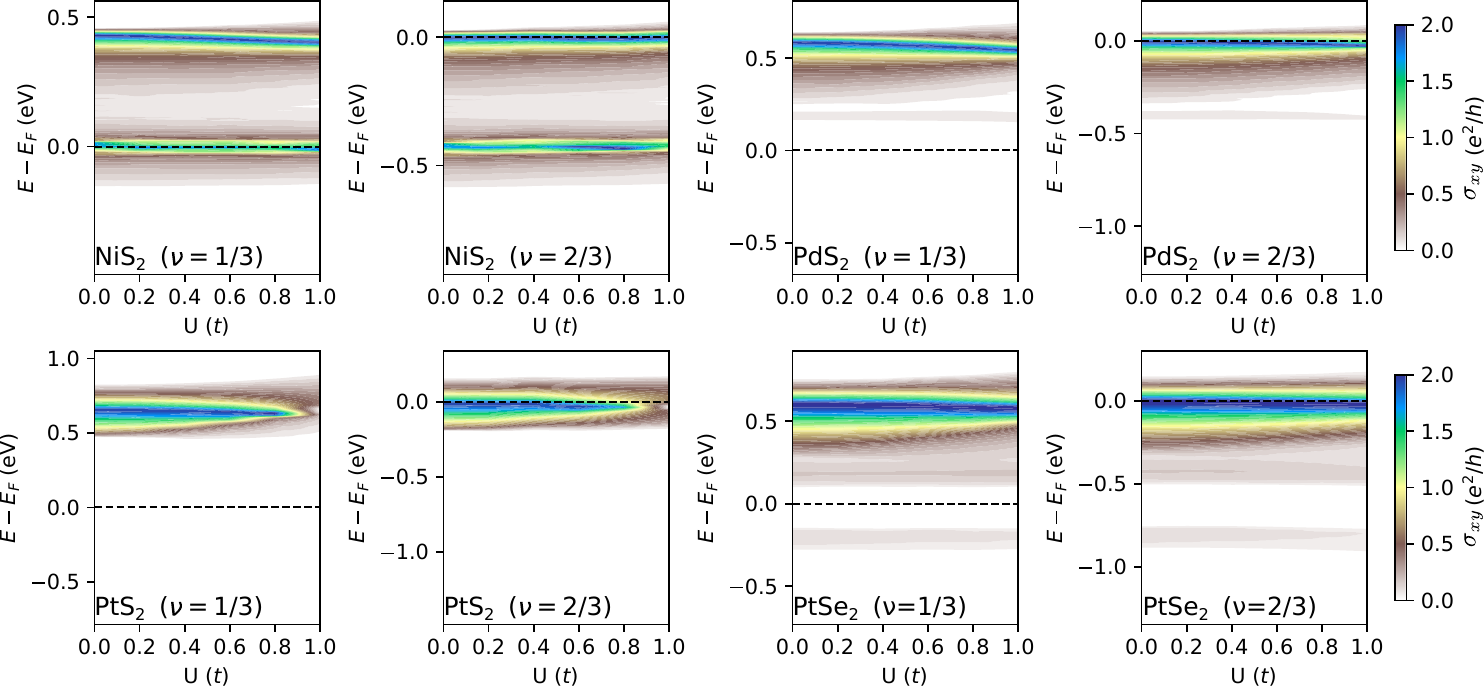}
    \caption{\label{fig:shc} 
    Spin Hall conductivity $\sigma_{xy}$ for different values of $U$ and electron occupations ($\nu=1/3$ and $\nu = 2/3$) in the MX$_2$ system. The color bars indicate the $\sigma_{xy}$ values.
    }
\end{figure*}

Inter vacancy interactions play a crucial role in driving the topologically phase in MX$_2$ systems.
Initially, considering $\lambda = 0.1\,t$, we observe the appearance of band dispersions and associated spin Hall conductivity calculated using the Kubo formula \cite{PRLsinova2004, PRBmatthes2016}, as a function of $\alpha$ [Figs.~\ref{fig:band-shc} (a1)--(b3)].
Here, the spin Hall conductivity is calculated considering the electron filling number fix ($\nu = 2/3$), and varying the range of energy integration up to $E - E_F$.
For lower vacancy densities (lower $\alpha$) the system exhibits a trivial virtual gap between the top-most bands, as depicted in Figs.~\ref{fig:band-shc}(a1)--(b1).
The upper gap transitions to a topological phase when the inter-vacancy interaction reaches a critical value $\alpha_c=(2-\beta) \lambda/ \sqrt{3}t$.
For $\beta=0$, $\alpha_c = 2\lambda/t\sqrt{3}$ defines the critical value for the virtual gap closing and reopening at $\nu=2/3$.
Specifically, for $\lambda = 0.1\,t$ the critical value is $\alpha = 0.11$, as evident from the gap closing in Figs.~\ref{fig:band-shc}(a2)--(b2).
For inter-vacancy interactions $\alpha > \alpha_c$, the system enters a non-trivial phase for $\nu=2/3$ filling, as illustrated in Figs.~\ref{fig:band-shc}(a3)--(b3).
Interestingly, as $\alpha \rightarrow 1$ the gap at $\nu = 1/3$ filling undergoes a topological phase transition.
In this limit, the system's orbitals become the kagome lattice.

\subsection{Electron-electron interaction}

Let us now consider the effect of electron-electron repulsion $U$ and electronic filling $\nu$ on the topological properties of our model system. 
Our calculations reveal that, in general, the topological non-trivial phase is governed by the interplay between inter-vacancy interaction $\alpha$ and spin-orbit coupling $\lambda$, with increasing $U$ degrading the non-trivial phase.
As a specific example, let us consider a non-interacting system in a topological phase for the $\nu =2/3$ filling, with parameters $\lambda = 0.1t$ and $\alpha = 0.3$.
The effect of $U$ on this topological phase is illustrated by the bottom row of Fig.~\ref{fig:band-shc}.
For $U$ up to $0.5t$ the system remains with a non-trivial topological phase characterized by $\sigma_{xy} = 2e^2/h$ [Figs.~\ref{fig:band-shc}(a4)--(b4)].
A topological phase transition occurs around $U=0.75t$, indicated by the closing of the gap [Figs.~\ref{fig:band-shc} (a5)--(b5)].
The gap reopens for larger $U$, [Figs.~\ref{fig:band-shc}(a6)--(b6)], being trivial with $\sigma_{xy}<2 e^2/h$. 

Figure~\ref{fig:phase} shows a topological phase diagram as a function of the electron-electron repulsion $U$ and the vacancy-vacancy interaction $\alpha$ for the top-most energy gap. 
Figure~\ref{fig:phase}(a) displays the ``virtual" 
topological phase diagram (at $\nu=1/3$), while Fig.~\ref{fig:phase}(b) shows the phase diagram for systems with the Fermi energy within the topological gap (at $\nu=2/3$). 
We first notice that a topological phase emerges only for $\alpha > \alpha_c$, a critical value dictated by the SOC strength $\lambda$. 
For any given $\alpha > \alpha_c$, increasing $U$ drives the system towards a trivial phase.
For more localized wave functions (lower $\alpha$ values), both in the virtual ($\nu=1/3$) and real ($\nu=2/3$) electron configuration exhibit similar phase diagrams.
However, at higher $\alpha$ values, the virtual topological gap proves more robust than the real topological gap.
The introduction of additional electrons enhances repulsion, further degrading the non-trivial phase.  
Figure~\ref{fig:phase}(c) illustrates the subtraction between the real topological gap phase diagram by the virtual phase diagram, defined as $\Delta {\cal C}_ s \equiv {\cal C}_s^{\nu=1/3} - {\cal C}_s^{\nu=2/3}$.
For $\alpha$ ranging from 0.1 to 0.4, the virtual topological phase is more robust against variation in $U$.
In summary, the phase transition is accompanied by a gap closure govern by the competition between vacancy density, SOC, and electron-electron repulsion satisfying the approximate relation, $E_{\rm gap} \approx [|a U^2 + \lambda^2 - (\alpha t\sqrt{3}/2)^2|]^{1/2}$.
The gap closure, $E_{\rm gap} =0$, is represented by the red curves in Figs.~\ref{fig:phase}(a) and (b), corresponding to the topological phase transition condition.
Here, as before, $a$ is a parameter that depends on $\nu$.
Figure~\ref{fig:phase}(d) presents a ternary diagram visualizing the topological phase behavior. 
Increasing the vacancy density stabilizes the non-trivial phase, while stronger SOC and electron-electron repulsion tend to drive the system towards a trivial phase.

\subsection{Topological robustness}

We proceed with realistic TMDs calculations that include vacancy states in the semiconducting gap, enabling the emergence of a topological insulator phase.
We identify such systems using density functional theory (DFT) calculations, with the extracted the corresponding tight-binding parameters for $H_{\rm intra}$ and $H_{\rm inter}$ [Table~\ref{tab:par-DFT}]. 
To address the uncertainties in the Hubbard $U$-value, particularly with electron doping, we extract our model parameters using $U=0$ DFT, effectively capturing the screened electronic interactions in the weak electronic correlations regime \cite{PRBschluter1989, PRLschluter1979}, and systematically vary $U$.
Given the nature of SOC, its strength is larger for heavier transition metals, ranging from $7\%$ to $18\%$ of the hopping strength.
The vacancy concentration dictates the proximity of adjacent vacancies and, consequently, the 
(second-neighbor) inter-vacancy decaying factors, $\alpha$ and $\beta$.
The later are obtained from DFT calculations for $n_{\rm vac} = 10^{14}$\,cm$^{-2}$ \cite{NATMATlin2017, 2DMlima2023}.
For such concentration, $\alpha$ ranges from $0.37$ for PtS$_2$ to nearly $1.00$ for NiS$_2$.
This variation in $\alpha$ reflects the localization of the defect wave function, being less localized in the NiS$_2$ system.
Due to the dielectric nature of the MX$_2$, the screening of the SOC field leads to $\beta \rightarrow 0$, validating our previous discussion.

\begin{table}[h!]
\caption{\label{tab:par-DFT} Model parameters extracted from DFT calculations \cite{NLlima2021} for  $n_{\rm vac} = 10^{14}$ cm$^{-2}$.}
\begin{ruledtabular}
    \begin{tabular}{c|cccc}
       MX$_2$  & $t$ (eV) & $\alpha$ & $\lambda$ (eV) & $\beta$  \\
       \hline
        NiS$_2$  & 0.15 & 0.99 & 0.01 & 0.03 \\
        PdS$_2$  & 0.26 & 0.58 & 0.02 &-0.05 \\
        PtS$_2$  & 0.36 & 0.37 & 0.06 &-0.10 \\
        PtSe$_2$ & 0.28 & 0.61 & 0.05 & 0.08   
    \end{tabular}
\end{ruledtabular}
\end{table}

Figure~\ref{fig:shc} shows the calculated spin Hall conductivity for the TMD systems [parameter in Table \ref{tab:par-DFT}] for different values of $U$ and both $\nu = 1/3$ and $\nu = 2/3$ fillings.
The robustness of the topological phase with respect to $U$ is direct correlated with the vacancy state localization.
Less localized systems, such as NiS$_2$, exhibit a topological phase  robust up to $U=\,t$, while in the more localized systems, PtS$_2$, the topological phase vanish for $U>0.6\,t$.
Comparing the robustness of the topological states at $\nu = 1/3$ filling with $\nu = 2/3$ fillings, we find that the more localized systems exhibit a virtual topological phase more robust than the real topological state. 
For instance, in PtS$_2$, the non-trivial phase degrades with lower $U$. 
These findings support our discussion of the topological phase diagrams shown in Fig.~\ref{fig:phase}.

\section{Conclusions}

Our study reveals that electron-electron interactions significantly impact the stability of topological phases, particularly the transition between virtual and real of topological states upon electron filling.  
As a case study, we demonstrated that in 2D TMDs with vacancy-induced topological phases, the robustness of these virtual/real phases is influenced by a complex interplay between vacancy density, spin-orbit coupling, and electron-electron repulsion. While higher vacancy concentrations can enhance the non-trivial topology, electron-electron interactions may suppress the topological phase, drastically reducing the number of possible experimental realizations of TIs.
These findings provide a comprehensive understanding on the interaction between electron filling, vacancy concentration, and topological protection in TMDs, highlighting the possibility of tuning topological phases in other materials through vacancy engineering and controlled doping.

\begin{acknowledgments}
The authors acknowledge financial support from the Brazilian agencies FAPESP (23/09820-2), CNPq (INCT - Materials Informatics), FAPERJ, and LNCC (Laboratório Nacional de Computação Científica) for computer time (Project ScafMat2 and DIDMat).
\end{acknowledgments}


\bibliography{bib}       

\end{document}